\documentclass[a4paper,11pt]{article}
\usepackage{pos}
\usepackage{graphicx}

\def\svev#1{\left\langle #1\right\rangle}       

\def\sl#1{\rlap{\hbox{$\mskip 1 mu /$}}#1}      

\newcommand{\bee}{\begin{equation}}
\newcommand{\ee}{\end{equation}}
\newcommand{\beea}{\begin{eqnarray}}
\newcommand{\eea}{\end{eqnarray}}

\title{Funny business from the large $N_c$ finite temperature crossover}
\ShortTitle{Funny business from the large $N_c$ finite temperature crossover}

\author*[a]{Thomas DeGrand}

\affiliation[a]{
Department of Physics,
University of Colorado, \\ Boulder, CO 80309, USA}

\emailAdd{thomas.degrand@colorado.edu}

\abstract{
It is  well known that the deconfinement transition temperature for $SU(N_c)$ gauge theory is almost independent of
$N_c$, and the transition is first order for $N_c \ge 3$. In the real world ($N_c=3$, light quarks) it
is a crossover located far away from the pure gauge value. What happens to the transition temperature at fixed fermion mass if 
the number of fermion flavors is held constant ($N_f=2$) and  $N_c$ is varied? There are multiple
plausible stories, only one of which appears to be true when the systems are simulated on the lattice.
I describe the physics issues which surround the question and my lattice - based answer to it.
}

\FullConference{%
 The 38th International Symposium on Lattice Field Theory, LATTICE2021
  26th-30th July, 2021
  Zoom/Gather@Massachusetts Institute of Technology
}

\begin{document}
\maketitle

The limit of QCD when the number of colors $N_c$ becomes large is a theorist's playground for
studying QCD  \cite{tHooft:1973alw,tHooft:1974pnl,Witten:1979kh}.
   It has a small lattice literature, checking its predictions for nonperturbative
quantities such as masses and matrix elements.
(See  Refs.~\cite{Lucini:2012gg,GarciaPerez:2020gnf,Hernandez:2020tbc} for reviews.)
In most cases with a lattice study, the lattice confirms its simple large $N_c$ prediction.

Why study large $N_c$ QCD with lattice methods? It's well known that QCD simplifies at large $N_c$
(it basically becomes a theory of open strings). QCD also idealizes at large $N_c$: mesons are
$q \bar q$ bound states, baryons are bound states of $N_c$ quarks, and hadronic wave functions
are presumed to become independent of $N_c$ as shown by the $N_c$ scaling of matrix elements.
Most large $N_c$ predictions are of nonperturbative quantities (even though they are often based
on color counting for Feynman diagrams) and these predictions ought to be subjected to nonperturbative tests.

Large $N_c$ lattice calculations are pretty straightforward: Simulate
across $N_c$ at fixed  bare 't Hooft coupling $\lambda=g^2 N_c$ or $\beta\propto N_c^2$.  Discover that 
this fixes the lattice spacing to be nearly equal across $N_c$.  Measure the same
observables across $N_c$. Scale the observable appropriately (for example, decay constants scale as $\sqrt{N_c}$)
and observe curve collapse (again for example $f_{PS}/\sqrt{N_c}$ versus quark mass). 

(Disclaimer: often people use some gluonic observable like the string tension or flow parameter to match lattice
spacings, rather than the bare coupling. The results are the same up to $1/N_c$ corrections, which can
be pushed from place to place by making different scale setting choices.)

Usually, there is a simple story for any observable at large $N_c$. But there is one observable which lacks a simple story 
-- or -- said better -- there are at least three simple stories. That is the finite temperature crossover for QCD with a fixed number
of flavors, as a function of the quark mass and $N_c$.

The first possibility comes from the naive large $N_c$ expectation that gluonic degrees of freedom dominate
fermionic ones in the large  $N_c$ limit.
The pure gauge transition is first order for $N_c\ge 3$ and the transition temperature is nearly
independent of $N_c$ \cite{Lucini:2005vg,Lucini:2012wq}.
 As the fermion mass falls from infinity the transition becomes a crossover.
Shouldn't the transition remain first order and at the same value as the pure gauge transition,
with the end point pushing to ever smaller fermion mass as $N_c$ rises?

The second scenario assumes the physics of the transition is dominated by naive chiral symmetry breaking.
 QCD at all $N_c$
has an $SU(N_f) \times SU(N_f)$ symmetry which 
undergoes spontaneous symmetry breaking to $SU(N_f)$. The Pisarski - Wilczek analysis
 \cite{Pisarski:1983ms} approximates the Goldstone sector
as a linear sigma model. For $N_f=2$ the system is expected to have a second order transition
at zero fermion mass, with $O(4)$ critical exponents. Second order transitions are unstable under
perturbation, so the transition becomes a crossover away from $m_q=0$.
This should be true for any $N_c$.

The issue then is,
how does the crossover temperature scale with $N_c$?
Linear sigma models contain one dimensionful parameter, the vacuum expectation value of the
scalar field, and all dimensionful observables (the pseudoscalar decay constant $f_{PS}$, and the
crossover temperature $T_c$ itself) are proportional to it. We already know that
$f_{PS} \propto\sqrt{N_c}$. Thus the naive prediction
of the second scenario is $T_c \propto \sqrt{N_c}$ \cite{Pisarski}.  

The third scenario is older than QCD.
 Confining theories are expected to show an exponentially growing spectrum of
resonances with mass, forming a Hagedorn spectrum  \cite{Hagedorn:1968zz}.
  The tower of resonances implies a limiting temperature $T_0$ and this implies
a crossover temperature $T_c \sim T_0$ \cite{Cabibbo:1975ig}. In the real world,
 the Hagedorn temperature is about 160 MeV.
The extension of the story for large $N_c$ and nonzero $N_f$ is that the spectrum of meson resonances
is basically identical across $N_c$.  If two
theories have the same spectrum, then they ought to
have the same critical properties.  
So the third prediction is that  any $N_c \ne 3$
with $N_f=2$ will qualitatively resemble $N_c=3$, $N_f=2$.
(Since large $N_c$ with nonzero $N_f$ is different from quenched QCD,
which only has glueballs, its Hagedorn temperature is different.)

This year I finished a little large $N_c$ study of  $N_f=2$ QCD with $N_c=3$, 4, and 5 with
medium heavy fermions, $(m_{PS}/m_V)^2 \sim 0.25-0.65$   \cite{DeGrand:2021zjw}.
The answer I found is that the third scenario is most correct:  the three systems show nearly identical crossover behavior 
as a function of temperature. 

All the technical details are in the paper, but in a few words, this is what I did. I used clover fermions.
The chiral condensate for Wilson type fermions has a set of divergent $1/a^n$ pieces, but the temperature dependent
condensate is well defined \cite{Borsanyi:2012uq,Borsanyi:2015waa}
\bee
\svev{\bar \psi \psi}_{sub} = \svev{\bar \psi \psi}_T - \svev{\bar \psi \psi}_{T=0}.
\ee
I measured
\bee
\frac{3}{N_c} t_0^{3/2} \Sigma(T) = \frac{3}{N_c} t_0^{3/2} \times m_q (\Delta_{PP}(T) - \Delta_{PP}(T=0) )
\label{eq:mycond}
\ee
where (explicitly showing the conversion from the lattice quantity computed from clover fermions to a continuum one)
\bee
\Delta_{PP}(T) = \hat \Delta_{PP}(N_t) (1-\frac{3\kappa}{4\kappa_c})^2 .
\ee
The lattice observable is
\bee
\hat \Delta_{PP}(N_t) = \sum_{t=0}^{N_t} \sum_x \svev{P(x,t) P(0,0)}   
\label{eq:delta}
\ee
where $P(x,t)=\bar \psi(x,t)\gamma_5 \psi(x,t)$ is the pseudoscalar current.
The factor of $t_0^{3/2}$ in Eq.~\ref{eq:mycond}
makes the observable dimensionless and the overall factor of $3/N_c$ is included to show
collapse to a common curve across
$N_c$ when the condensate scales proportional to $N_c$ as expected by large $N_c$ counting.

\begin{figure}
\begin{center}
\includegraphics[width=0.9\textwidth,clip]{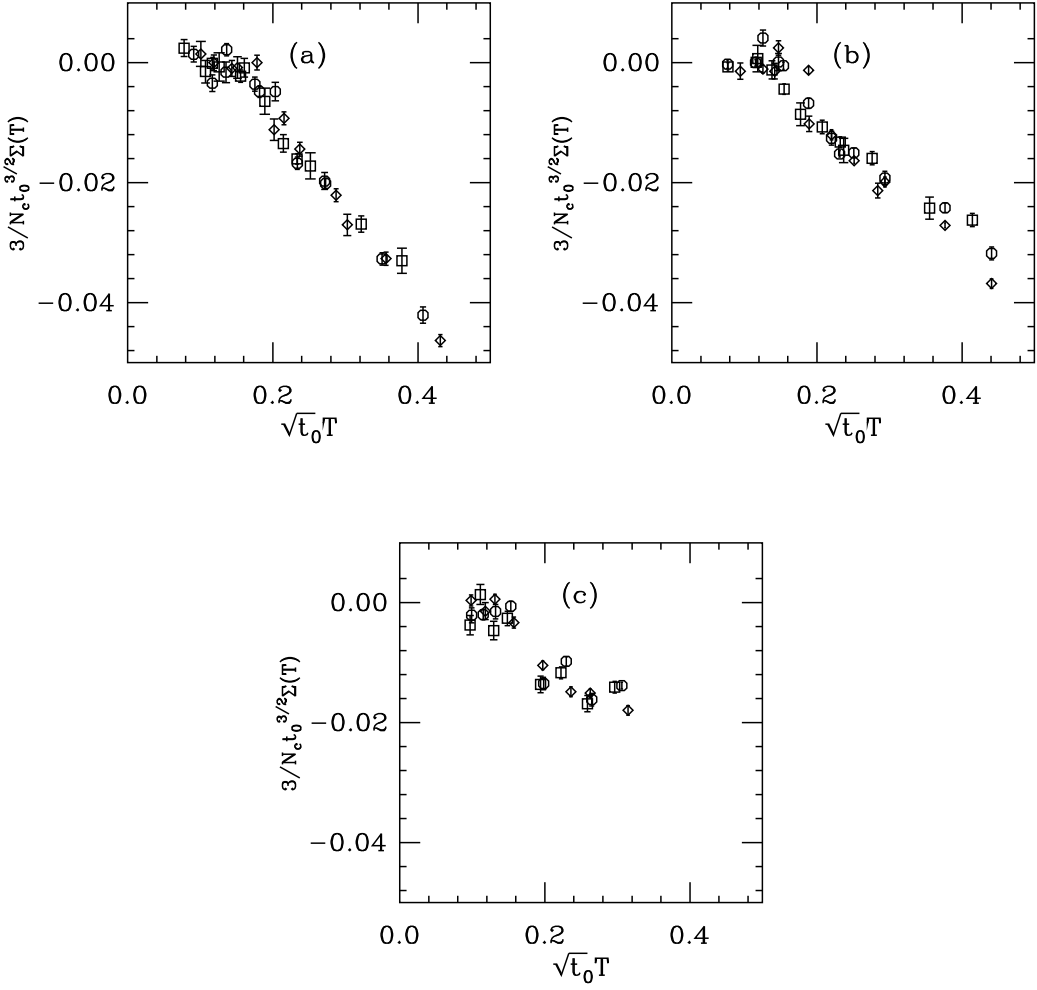}
\end{center}
\caption{The temperature dependent condensate, rescaled by $3/N_c$,
as a function of temperature, in appropriate units of $t_0$.
Squares, octagons, and diamonds label $N_c=3$, 4, and 5.
(a)  $(m_{PS}/m_V)^2 \sim 0.63$;
(b) $(m_{PS}/m_V)^2 \sim 0.5$;
(c) $(m_{PS}/m_V)^2 \sim 0.25$.
\label{fig:sigmaall}}
\end{figure}

\begin{figure}
\begin{center}
\includegraphics[width=0.9\textwidth,clip]{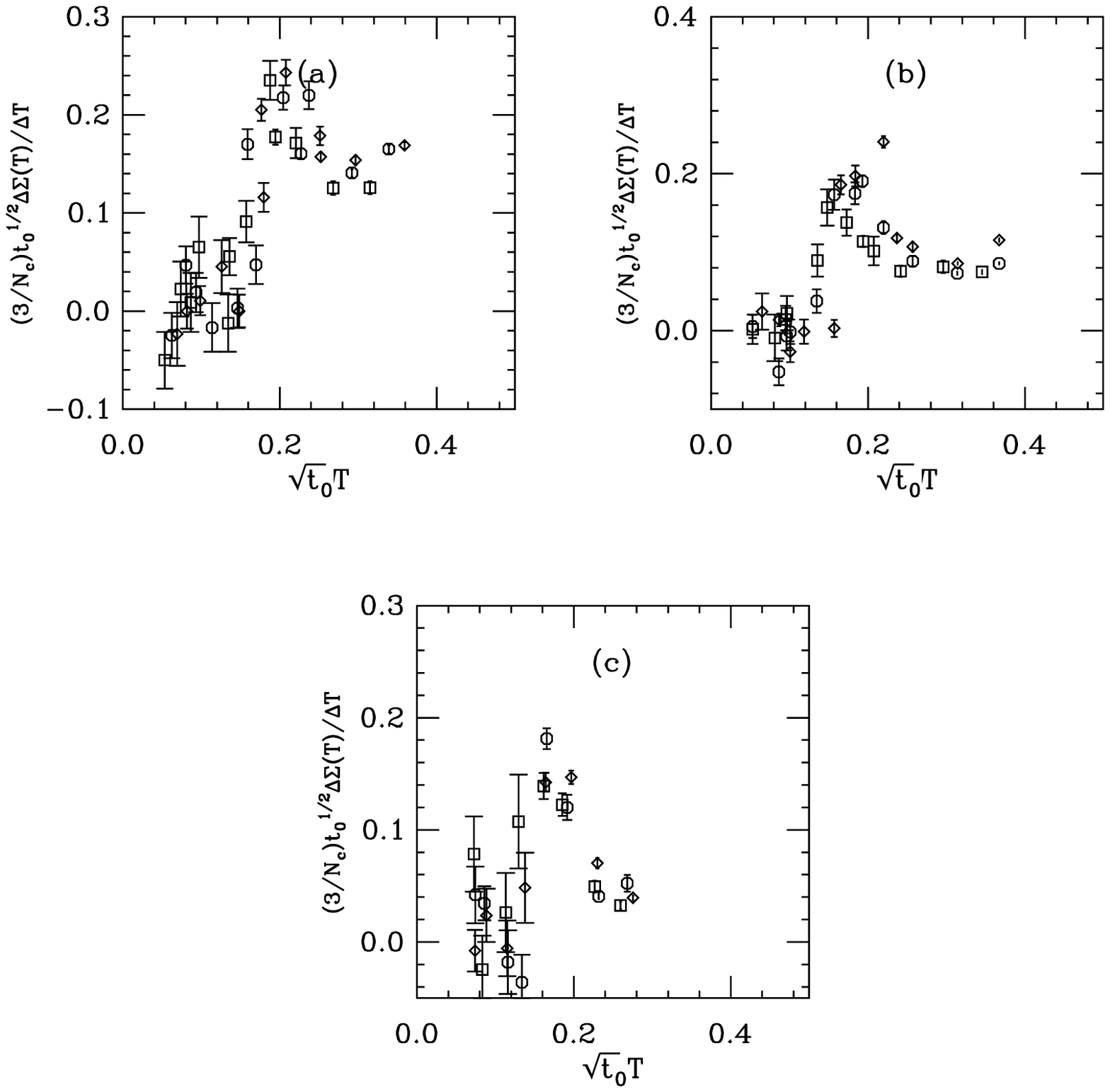}
\end{center}
\caption{$\Delta \Sigma(T)/\Delta T$, rescaled by $3/N_c$,
as a function of temperature, in appropriate units of $t_0$.
Squares, octagons, and diamonds label $N_c=3$, 4, and 5.
(a)  $(m_{PS}/m_V)^2 \sim 0.63$;
(b) $(m_{PS}/m_V)^2 \sim 0.5$;
(c) $(m_{PS}/m_V)^2 \sim 0.25$.
\label{fig:slopeall}}
\end{figure}

Two figures from the paper show the results: the condensate itself in Fig.~\ref{fig:sigmaall} and 
$\Delta \Sigma(T)/\Delta T$ in Fig.~\ref{fig:slopeall}.

How dull can things be? There is a smooth crossover showing chiral restoration at the same temperature across $N_c$.
The derivative in Fig.~\ref{fig:slopeall} was my attempt to see something with a peak. There is a peak but it is very broad.
The location of the crossover is intermediate between the pure gauge result of Refs.~\cite{Lucini:2005vg,Lucini:2012wq}
and $SU(3)$ results at the physical point. There is no sign of anything first order.

So what can we make of this? There is no first order transition. Apparently, fermions are still sometimes important degrees of freedom at large $N_c$.

What about Pisarski - Wilczek?
It's  well known that pions in QCD are NOT described by a linear sigma model (see Gasser and Leutwyler, Ref.~\cite{Gasser:1983yg}).
The linear sigma model is really only used to make statements about the critical properties of a system. This does not
include the critical temperature, which is not universal. Only critical exponents are universal.

What about a connection with the spectrum? It's not possible to test for a Hagedorn spectrum directly (yet) for any $N_c$.
(Perhaps one should be more modern and talk about the hadron resonance gas model -- a sum over
all the resonances in the Review of Particle Properties.)
 But basically all mesons created with an interpolating field $\bar \psi \Gamma \psi$ mesons (S-wave
and P-wave mesons) are known to have an $N_c-$independent spectrum. Furthermore, the scaling of the pseudoscalar decay constant
 $f_{PS}  \propto \sqrt{N_c}$ means that  the amplitude for pion scattering $A(\pi \pi\rightarrow \pi\pi) \propto 1/N_c$:
pions (like other hadrons) don't interact at large $N_c$. In addition,
large $N_c$ scaling says  that the vector meson decay constant $f_V  \sim \svev{\gamma|\rho}\propto \sqrt{N_c}$ and vector dominance
says $g_{\rho\pi\pi} \propto  1/\svev{\gamma|\rho} \propto 1/\sqrt{N_c}$. (For a textbook discussion, see Ref.~\cite{Feynman:1973xc},
being careful with conventions defining $f_V$.)
 Again, hadrons don't interact at large $N_c$.
So what else is left but the density of states?

There is a lattice test one could do: compare the
 trace anomaly  $(\epsilon - 3P)/T^4$ for $T<T_c$ ($\epsilon$ is the energy density, $P$ is the pressure) 
  to a hadron resonance gas model, or just compare one $N_c$ to another. Results from all $N_c$'s should match.

Let's summarize:

My numerics were not very high quality but the effect was so obvious, it didn't matter.

It might be interesting to do large $N_c$ thermodynamics ``right.'' This probably means using staggered fermions since the volume scaling for
thermodynamics is so fierce. A lot of the continuum quark gluon plasma phenomenology  is large $N_c$ based (AdS/CFT certainly is)
and some predictions might be checked by going to large $N_c$ on  the lattice.


\begin{acknowledgments}
I am grateful to Rob Pisarski for a conversation about large $N_c$ expectations for QCD thermodynamics.
My computer code is based on the publicly available package of the
 MILC collaboration~\cite{MILC}. The version I use was originally developed by Y.~Shamir and
 B.~Svetitsky.
This material is based upon work supported by the U.S. Department of Energy, Office of Science, Office of
High Energy Physics under Award Number DE-SC-0010005.
Some of the computations for this work were also carried out with resources provided by the USQCD
Collaboration, which is funded
by the Office of Science of the U.S.\ Department of Energy
using the resources of the Fermi National Accelerator Laboratory (Fermilab), a U.S.
Department of Energy, Office of Science, HEP User Facility. Fermilab is managed by
 Fermi Research Alliance, LLC (FRA), acting under Contract No. DE- AC02-07CH11359.
\end{acknowledgments}



\end{document}